\documentclass[11pt]{article}

\usepackage{lineno}

\modulolinenumbers[1]

\usepackage{pstricks}

\newcommand{\lb}{\left(}
\newcommand{\rb}{\right)}

\usepackage{amsmath}
\usepackage{amssymb}
\usepackage{color}
\usepackage{soul}
\usepackage{graphicx}

\usepackage{setspace}

\begin{document} 


\begin{center}
{\LARGE {\bf Extended landslide velocity and analytical drag}}
\\[7mm]
{\large {Shiva P. Pudasaini}}\\[2mm]
Technical University of Munich\\
School of Engineering and Design, Chair of Landslide Research\\
{Arcisstrasse 21, D-80333, Munich, Germany}\\[1mm]
{E-mail: shiva.pudasaini@tum.de}\\[8mm]
\end{center}
{\bf Abstract:} 
The landslide velocity plays a dominant role in estimating the impact force and devastated area. Here, based on Pudasaini and Krautblatter (2022), 
I develop a novel extended landslide velocity model that includes the force induced by the hydraulic pressure gradient, which was neglected by all the existing analytical landslide velocity models. By a rigorous conversion between this force and inertia, which facilitates constructing exact analytical solutions for velocity, I develop two peer systems expecting to produce the same results. However, this contradicts with our conventional wisdom. This raises a legitimate question of whether we should develop some new balance equations such that these phenomena can be better explained naturally. I compare the two velocity models that neglects and includes the force induced by the hydraulic pressure gradient. Analytical solutions produced by the two systems are fundamentally different. The new model is comprehensive, elegant, and yet an extraordinary development as it reveals serendipitous circumstances \mbox{resulting} in a \mbox{pressure-inertia-paradox}. Surprisingly, the mass first moves upstream for quite a while, then it winds back and continues accelerating down slope. The difference between the extended and simple solution is significant, and widens strongly as the force associated with the hydraulic pressure gradient increases, demonstrating the importance of this force in the landslide velocity. The viscous drag is an essential dissipative force mechanism and plays an important role in controlling the landslide dynamics. However, no explicit mechanical and analytical model exists to date for this. The careful sagacity of the graceful form of new velocity equation results in a plain, yet mechanically extensive, analytical model for viscous drag, the first of this kind. It contains several physical and geometrical parameters, and evolves dynamically as it varies inversely with the flow depth. A dimensionless drag number is constructed characterizing the drag dynamics. Importance of the drag model is explained. In contrast to the prevailing practices, I have proven that drags are essentially different for the expanding and contracting motions. This is an entirely novel perception. Drag coefficients are close to the empirical or numerical values often used in practice. But, now, I offer an innovative, physically-founded analytical model for the drag that can be instantly applied in mass flow simulations.   

\section{Introduction}

Mountain torrents and valleys are often hit by landslides and debris flows threatening the lives, devastating civil properties, and eternally swapping the landscape (Evans et al., 2009; Mergili et al., 2020, Shugar et al., 2021). Adequate understanding of the dynamics, runout and deposition processes of mass transports still remains a great challenge to the mountain engineers, practitioners, flow modelers and the hazard assessment professionals (Cuomo et al., 2016; Dietrich and Krautblatter, 2019). The dynamics of landslide and the associated hazard is directly related to, and primarily controlled by, its velocity. So, the landslide velocity plays a fundamental role for the assessment of landslide hazards, mitigation measures, design of protective structures and landuse planning (Johannesson et al., 2009; Dowling and Santi, 2014; Faug, 2015). However, our understanding of the mechanical controls of evolving velocity, runout and impact energy of the landslide is still poor. Appropriate modelling of landslide impact force demands for a comprehensive knowledge of its velocity.
\\[3mm]
There are three major ways of studding the landslide dynamics: field or laboratory investigations and experiments, physical-mathematical modelling and numerical simulations, and analytical solutions (Pudasaini and Hutter, 2007). The available field data on landslides play important role in acquiring basic information on their dynamics. However, such data are insufficient and cannot cover the broad spectrum. The properties of the landslide material and its interaction with the boundary are difficult to determine to a high degree of accuracy. The dynamic field data are scarce and the aftermath static data are often only available for single locations. Much of the low resolution measurements are locally or discretely based on points in time and space (McCoy et al., 2012; Theule et al., 2015; de Haas et al., 2020). So, the proper understanding and interpretation of the field data pose challenges. The well controlled and desirably obtained data from laboratory or field experiments (Iverson and Ouyang, 2015; de Haas and van Woerkom, 2016; Baselt et al., 2021) can overcame many of the problems associated with the field events. But, still such data cannot cover the whole picture of the dynamics and deposition because of their largely reduced scales, substantially simplified materials and flow paths, and neglection of many of the important interactions between the involved phases. Other fundamental problems are also directly associated with the complex interactions of the natural events with their boundaries, which cannot be fully understood by controlled experiments.
\\[3mm]
Due to its overall nature and capabilities of covering the complex materials and their evolutions, interactions with the environment, applicabilities to any natural slopes and torrents, the theoretical modelling (Le and Pitman, 2009; Pudasaini, 2012; Pudasaini and Mergili, 2019) have gain a tremendous pace in the past decades, and have become the major solutions of the problems associated with the mass flow dynamics. A number of comprehensive numerical modelling for mass transports are available (McDougall and Hungr, 2005; Christen et al., 2010; Frank et al., 2015; Iverson and Ouyang, 2015; Cuomo et al., 2016; Mergili et al., 2020; Liu et al. 2021). Yet, numerical simulations are approximations of the physical-mathematical model equations and their validity is often evaluated empirically (Mergili et al., 2020). In contrast, exact, analytical solutions can provide better insights into the complex flow behaviors, and provide the fastest, the most cost-effective and the best rigorous answer to the problem (Gauer, 2018; Faraoni, 2022; Pudasaini and Krautblatter, 2022). Moreover, exact solutions to non-linear model equations are necessary to elevate the accuracy of numerical solution methods based on complex numerical schemes (Chalfen and Niemiec, 1986; Pudasaini and Krautblatter, 2022). This is very useful to interpret complicated simulations and avoid mistakes in numerical simulations. However, the numerical solutions (Mergili et al., 2020; Shugar et al., 2021) can cover the widest spectrum of the complex flow dynamics described by advanced mass flow models (Pudasaini and Mergili, 2019). Once such simulations are tested and validated against the analytical solutions, they may provide even more accurate results than the simplified analytical solutions.
\\[3mm]
Since the middle of the 20th century, a number of analytical solutions have been presented for landslides, avalanches and debris flows (Voellmy, 1955; Salm, 1966; Perla et al., 1980; McClung, 1983). The existing analytical velocity models and their solutions are effectively valid to the mass point or center of mass motion, and did not consider internal deformation. Pudasaini and Krautblatter (2022) overcame those deficiencies by introducing a simple, physics-based general analytical landslide velocity model that simultaneously incorporates the internal deformation and externally applied forces, consisting of the net driving force and the viscous resistant. They showed that the non-linear advection and external forcing fundamentally regulate the state of motion and deformation. They constructed several general exact analytical solutions. Those solutions cover the wider spectrum of landslide velocity and directly reduce to the mass point motion. 
\\[3mm]
Here, following Pudasaini and Krautblatter (2022), I develop a novel extended landslide velocity model that includes the lateral force induced by the hydraulic pressure gradient, which, however, was neglected by all the existing analytical landslide velocity models. By a fully consistent conversion between this force and inertia, I develop two peer systems. Surprisingly, the solutions produced by these two systems appear to be fundamentally different. The difference widens strongly as the force associated with the hydraulic pressure gradient increases. The new extended landslide velocity model describes some unexpected situations emerging for the first time. The mechanism of the new solution is explained based on the contribution from the hydraulic pressure gradient. I also construct a simple, physically-founded distinctive analytical model for the viscous drag. A new dimensionless drag number is formulated. I prove that drags are essentially different for the expanding and contracting motions. The exact, analytical solutions derived in this paper may find applications in environmental, engineering and industrial mass transports down slopes and channels where the force induced by the hydraulic pressure gradient is significant which is the realistic situation. Similarly, the analytical drag model may serve the purpose of the practitioners and engineers involved in mass flow simulations and hazard mitigation.

\section{The model development}

A geometrically two-dimensional motion down a slope is considered. Let $t$ be time, $(x,
z)$ be the coordinates and $\left ( g^x, g^z\right )$ the gravity accelerations along and perpendicular to the slope, respectively. Let, $h$ and $u$ be the flow depth and the mean flow velocity of the landslide along the slope. Similarly,
$\gamma, \alpha_s, \mu$ be the density ratio between the fluid and the solid
particles $\left ( \gamma = \rho_f/\rho_s\right )$, volume fraction of the solid
particles (coarse and fine solid particles), and the basal friction coefficient $\left ( \mu =
\tan\delta\right )$, where $\delta$ is the basal friction angle of the solid particles, in
the mixture material. Furthermore, $K$ is the earth pressure coefficient, and $\beta$ is the viscous drag coefficient. Then, consider the mass and momentum balance equations describing the motion of a landslide (Pudasaini and Krautblatter, 2022):
\begin{equation}
\frac{\partial h}{\partial t} +  \frac{\partial }{\partial x}\left ( hu\right ) = 0,
\label{Eqn_1}
\end{equation}
\begin{equation}
\frac{\partial u}{\partial t} + u \frac{\partial u}{\partial x} 
+ \lambda \frac{\partial h}{\partial x}
= \alpha - 
\beta u^2,
\label{Eqn_2}
\end{equation}
where
\begin{equation}
\alpha: = g^x – (1-\gamma)\alpha_s g^z\mu, 
\label{Eqn_2al}
\end{equation}
is the net driving force (this includes the forces due to gravity, Coulomb friction, lubrication, and liquefaction) and $\beta$ is the viscous drag coefficient. So, $\alpha$ and $\beta$  constitute the net driving and the resisting forces in the system. Moreover, 
\begin{equation}
\lambda: = g^z\left \{ \left( \left( 1-\gamma\right)K + \gamma\right)\alpha_s+\left ( 1-\alpha_s \right )\right \}, 
\label{Eqn_2la}
\end{equation}
is the effective hydraulic pressure parameter. This is (associated with) the lateral load, and combined with the free-surface gradient ($\partial h/\partial x$), constitutes the force due to the hydraulic pressure gradient ($\lambda \partial h/\partial x$).
\\[3mm]
Assuming the steady-state ($\partial u/\partial t = 0$), equations (\ref{Eqn_1}) and (\ref{Eqn_2}) can, respectively, be written as
\begin{equation}
\frac{\partial h}{\partial x} = -\frac{h}{u}\frac{\partial u}{\partial x},
\label{Eqn_1a}
\end{equation}
\begin{equation}
u \frac{\partial u}{\partial x} 
= - \lambda \frac{\partial h}{\partial x}
 +\alpha - 
\beta u^2.
\label{Eqn_2a}
\end{equation}
 As we will see below, it is crucial to carefully analyze (\ref{Eqn_2a}) in relation with (\ref{Eqn_1a}). For the landslide accelerating downslope, both the parameters $\alpha$ and $\lambda$ are positive, so is the total system force, $\alpha - \beta u^2$. As the landslide accelerates down the slope, the velocity increases and its surface gradient decreases, i.e., $\partial u/\partial x > 0$, $\partial h/\partial x < 0$. Moreover, the velocity and depth of the landslide are positive quantities. Then, the term $-\lambda \partial h/\partial x$ in (\ref{Eqn_2a}), which stems from the hydraulic pressure gradient, is a positive force that enhances the motion. This means, all the force components on the right hand side of (\ref{Eqn_2a}) are positive, and consequently, $\left (-\lambda \partial h/\partial x +\alpha - 
\beta u^2 \right )> 0$. The left hand side of (\ref{Eqn_2a}) is the inertia of the system. The advective term, $u \partial u/\partial x$, tells that the landslide propagates downslope with the advection velocity $u$. And, this advection is caused by the overall composite forcing $\left (-\lambda \partial h/\partial x +\alpha - 
\beta u^2 \right )$. So, the positive force pulls the landslide along the slope in the positive $x$-direction, as expected. 
\\[3mm]
Dynamically, the left and right sides of (\ref{Eqn_1a}) are equivalent as it satisfies the mass balance. 
 Now, combining the momentum balance equation (\ref{Eqn_2a}) with the mass balance equation (\ref{Eqn_1a}), I obtain:
\begin{eqnarray}
u \frac{\partial u}{\partial x} - \lambda\frac{h}{u} \frac{\partial u}{\partial x}
=  \alpha - \beta u^2.
\label{Eqn_2b}
\end{eqnarray}
This can be written as 
\begin{eqnarray}
\left (u  - \frac{\lambda h}{u} \right)\frac{\partial u}{\partial x}
=  \alpha - \beta u^2,
\label{Eqn_2c}
\end{eqnarray}
which describes the landslide motion with the advection velocity 
 $\displaystyle{\left (u - \frac{\lambda h}{u}\right )}$.
\\[3mm]
{\bf The pressure-inertia-paradox:} As convenient, I analyze the landslide dynamics with respect to (\ref{Eqn_2b}) or (\ref{Eqn_2c}). It is clear that as compared with the inertia $u \partial u/\partial x$ in (\ref{Eqn_2a}), the inertia $\left ( u - \lambda h/u\right)\partial u/\partial x$ in (\ref{Eqn_2c}) is decreased, because $\lambda h/u >0$.
In principle, the systems represented by (\ref{Eqn_2a}) and (\ref{Eqn_2b}) should be equivalent, because there is a rigorous and fully consistent conversion between the force and inertia. In (\ref{Eqn_2a}), $\left (-\lambda \partial h/\partial x\right)$ acts as a force contribution induced by the hydraulic pressure gradient, which is positive for the accelerating landslide, so it enhances the motion. This means, the landslide described by the system (\ref{Eqn_2a}) accelerates downslope with the higher rate when $-\lambda \partial h/\partial x$ is present than when it is absent. In (\ref{Eqn_2b}), $-\left (\lambda h/u\right) \partial u/\partial x$ acts as an inertial contribution, and due to its negative sign, this also enhances the motion and the mass should accelerate downslope with the higher rate when $-\left ( \lambda h/u\right )\partial u/\partial x$ is present than when it is absent. 
Dynamically, increased force is equivalent to consistently decreased inertia. Consequently, the two peer systems (\ref{Eqn_2a}) and (\ref{Eqn_2b}) are expected to produce the same results. However, this does not appear to be happening here. To prove this, consider an initially relatively small velocity $u_0 > 0$. This is a legitimate choice, as often landslide is triggered with zero or small positive initial velocity. For the accelerating system under consideration, the total system force in (\ref{Eqn_2c}) is positive, i.e., $\left (\alpha - \beta u^2\right) > 0$. Which says that the mass must slide and accelerate down the slope. However, the inertial term on the left hand side of (\ref{Eqn_2c}) tells that, as long as $u_0 < \sqrt{\lambda h}$, the mass advects in the negative direction, means, it accelerates upslope for a certain time and distance. This contradicts with our conventional wisdom that the mass must accelerate downslope. 
This is a paradox, which I call the \mbox{pressure-inertia-paradox, $P_{pi}$}.
This is a novel understanding. Nevertheless, if we begin with the initial velocity such that $u_0 > \sqrt{\lambda h}$, then the landslide described by the system (\ref{Eqn_2c}) also accelerates downslope as that described by the system (\ref{Eqn_2a}). Yet, the solutions produced by these two systems are fundamentally different.  The results are presented in Section 3.1. 
\\[3mm]
As combining momentum balance with the mass balance is a common practice in fluid mechanics, I have proven that, one must very carefully analyze the dynamics of the resulting system, and check if it is consistent with the common perception. This is important. Note that, in principle, the solutions and results presented here are valid to any dynamical systems that can be structurally cast in the form (\ref{Eqn_2c}), irrespective of their specific physical parameters and meanings. 

\section{Extended landslide velocity model}

For ease of analysis, I write (\ref{Eqn_2c}) as 
\begin{eqnarray}
\lb u^2 - \lambda h\rb\frac{1}{u}\frac{\partial u}{\partial x} = \alpha - \beta u^2.
\label{Eqn_3}
\end{eqnarray}
This is a simple, yet very interesting equation, that characterizes some key and abnormal aspects of the flow dynamics. This stems from the elegant form of the new model. It is important to perceive that (\ref{Eqn_3}) implicitly includes the dynamics associated with the hydraulic pressure gradient via the mass balance. It also explicitly includes the depth of the landslide, $h$. The model (\ref{Eqn_3}) can be analyzed differently. I explore this in detail.
\\[3mm]
Equation (\ref{Eqn_3}) extends the Pudasaini and Krautblatter (2022) velocity model in which the hydraulic pressure gradient $\partial h/\partial x = h_g$ was assumed, and the term $\lambda\partial h/\partial x = \lambda h_g $ was combined with $\alpha$ to define the effective driving force $\alpha^e: = \alpha - \lambda h_g$. This means, Pudasaini and Krautblatter (2022) ignored the dynamic contribution of the hydraulic pressure gradient, but only parameterized it. 
 Moreover, all the existing analytical landslide velocity models do not consider the dynamics of the hydraulic pressure gradient.
For this reason, I call (\ref{Eqn_3}) the extended landslide velocity model. Assume that the flow depth $h$ can be parameterized with some representative value, say, typically the initial depth of the landslide, and set $\Lambda = \lambda h$, which has the dimension of the square of velocity. Then, (\ref{Eqn_3}) takes the form:
\begin{eqnarray}
\frac{\partial u}{\partial x} 
= \left [ \frac{u\left(\alpha - \beta u^2\right )}{u^2 - \Lambda}  \right ].
\label{Eqn_4}
\end{eqnarray}
This equation can be solved analytically to obtain the exact solution:
\begin{eqnarray}
\displaystyle{2\alpha \beta \left( x - x_0\right)=
\ln\left [ \left ( \frac{u_0}{u}\right )^{\left (2 \beta\Lambda\right)}\left( \frac{\alpha - \beta u_0^2}{\alpha - \beta u^2}\right)^{\left( \alpha - \beta \Lambda\right)} \right ]},
\label{Eqn_5}
\end{eqnarray}
where, $u_0 = u(x_0)$ is the initial (or boundary) condition at a given channel location $x = x_0$. 
The model (\ref{Eqn_3}), and (\ref{Eqn_4}) as well as its solution (\ref{Eqn_5}) are new.
As $\Lambda \to 0$ (this means: very thin flow depth, or almost vanishing free-surface gradient, or neglection of the hydraulic pressure gradient), (\ref{Eqn_5}) reduces to the simpler solution to the landslide velocity model by Pudasaini and Krautblatter (2022):
\begin{equation}
 u = \sqrt{\frac{\alpha}{\beta}\left[ 1 - \left ( 1- \frac{\beta}{\alpha}u_0^2\right)\frac{1}{\exp(2\beta(x-x_0))} \right]}.
\label{Eqn_6}
\end{equation}
 When $\Lambda \neq 0$, the solution (\ref{Eqn_5}) becomes physically more interesting and intricate. Depending on the values of $\Lambda$ and $u_0$, it describes different dynamics. 
  These aspects can now be explained by a single unified exact analytical solution in (\ref{Eqn_5}). Whereas the extended landslide velocity model and its solution (\ref{Eqn_5}) include the dynamically important force associated with the free-surface gradient, or the hydraulic pressure gradient, and the flow depth itself, the previous simple model and its solution (\ref{Eqn_6}) ignored this crucial feature. This is the novel aspect of the landslide velocity model (\ref{Eqn_3}). 
  
  \subsection{Results on extended landslide velocity}
  
  I consider plausible physical parameters (Pudasaini and Krautblatter, 2022) as: the slope angle $\zeta = 50^\circ$, basal friction angle $\delta = 20^\circ$, the fluid to solid density ratio $\gamma = 1100/2700$, solid volume fraction $\alpha_s = 0.65$, and the earth pressure coefficient $K = 1.0$ (extensional motion, Pudasaini and Hutter, 2007). This implies $\alpha = 6.6$, and $\lambda = 6.3$. Similarly, consider $\beta = 0.0019$.
  The results are shown in Fig. \ref{Fig_1} for initial conditions $x_0 = 0$, $u_0 = 0.2$, and different landslide depths $h = 2, 4, 6$. As $\Lambda = \lambda h$, the corresponding values for $\Lambda$ are $\Lambda = 12.61, 25.22, 37.80$. And, $\Lambda = 0$ indicates the simple solution given by (\ref{Eqn_6}), which does not include the effect of the flow depth $h$ and the hydraulic pressure gradient. Figure \ref{Fig_1} reveals very special aspects of landslide dynamics. The important observation is that for any (positive) value of the flow depth $h$, the mass first moves backwards for quite a while, then it winds back and accelerates down slope. This feature intensifies rapidly as $h$ increases. The model (\ref{Eqn_3}) and its solution (\ref{Eqn_5}) describe two very peculiar situations, which I examine below.
 
 \subsubsection{The upwind solution}
 
 For relatively small values of $u$ such that $\left (u - \Lambda/u\right) < 0$, the mass first advects in the upstream (negative) direction. This means, when the upstream dynamic wave celerity is negative $u - \sqrt{\Lambda} < 0$, or the flow is sub-critical, $Fr = u/\sqrt{\Lambda} < 1$, where $Fr$ is the Froude number, mass propagates upstream. However, due to the consideration of the mixture as composed of the solid particles and the fluid, the Froude number here is more general than the usual Froude number defined for the fluid or the dry grain flows (Pudasaini, 2012). 
  \begin{figure}[t!]
\begin{center}
  \includegraphics[width=15cm]{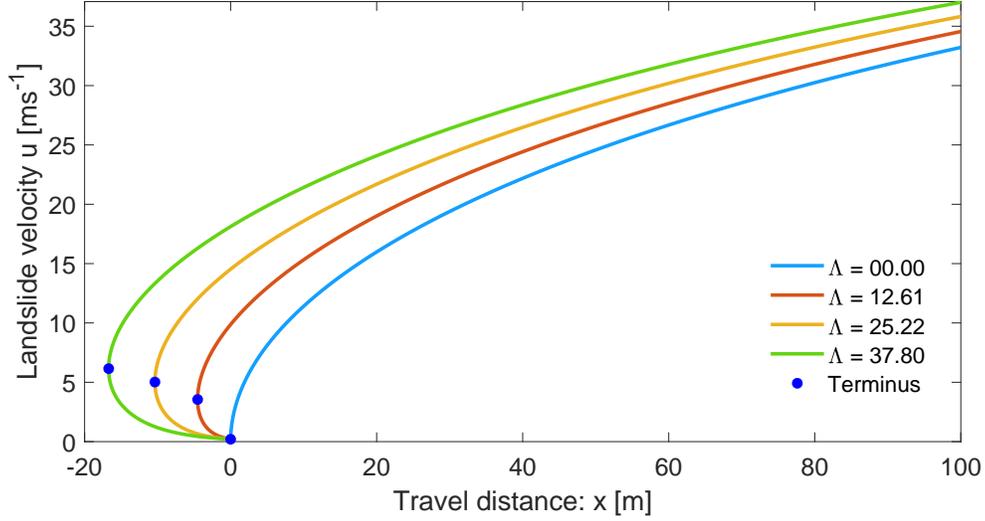}
  \end{center}
  \caption[]{Evolution of the landslide velocity as given by the extended landslide velocity model (\ref{Eqn_3}) and its analytical solution (\ref{Eqn_5}). The solutions for $\Lambda > 0$ include both the upwind branch and the downwind branch. The symbols {\blue $\bullet$} indicate the terminus (turning points) of the upwind branches and beginning of the downwind branches at (0,0), (-4.52, 3.55), (-10.37, 5.02), (-16.70, 6.15) as $\Lambda$ increases, whose values are indicated in the legend. For $\Lambda = 0$, there exists only the downwind branch.}
  \label{Fig_1}
  \end{figure}
 In order to facilitate the discussion on the landslide dynamics, consider $\Lambda = 37.80$ corresponding to $h = 6$.
 For this, the landslide accelerates upstream as far as $x_t = -16.7$ m with its maximum velocity of about $u_t= 6.15$ ms$^{-1}$, where $_t$ stands for terminus as shown in Fig. \ref{Fig_1} with the symbol {$\bullet$}. I call this the upwind branch of the solution. 
 This is the situation of the pressure-inertia-paradox, $P_{pi}$.
 Although mathematically rigorously derived solution, we still have to find likely scenarios for this branch of solution to be representative of a physical or natural situation. However, there can be something extra than our current understanding. We might even need to develop more general balance equations such that these phenomena can be better explained naturally.  
 
 \subsubsection{The downwind solution}
 
 In another situation, for relatively large values of $u$ such that $\left (u - \Lambda/u\right) > 0$, the mass advects in the downstream (positive) direction. This means, when the dynamic wave celerity is positive ($u - \sqrt{\Lambda} > 0$), or the flow is super-critical ($Fr = u/\sqrt{\Lambda} > 1$), the mass propagates downstream. With the above considered parameters, as the landslide attains its upstream maximum value of $u_t= 6.15$ ms$^{-1}$ at $x_t = -16.7$ m, then it winds down (come back), as indicated by terminus in Fig. \ref{Fig_1}, to the downstream and continues to accelerate along the down slope. I call this the downwind branch of the solution, which can be physically easily explained as usual for any landslide that is triggered with the initial velocity of $u_0 \ge \sqrt{\Lambda}$. This can be a scenario for  landslides triggered by a strong seismic shaking or a volcanic eruption. Or, this solution can be applied by considering the initial velocity at a location such that the condition $u_0 \ge \sqrt{\Lambda}$ is satisfied.  
 
  \subsubsection{The mechanism of solution (\ref{Eqn_5})}
  
  A closer look at the mathematical structure of the extended solution (\ref{Eqn_5}) reveals its intrinsic mechanism and the associated dynamics. There are two groups of terms internal to (\ref{Eqn_5}) inside the logarithm function $\ln$. These are: ($i$) $\displaystyle{\left (\frac{u_0}{u}\right)^{2 \beta \Lambda}}$, and ($ii$) $\displaystyle{\left( \frac{\alpha - \beta u_0^2}{\alpha - \beta u^2}\right)^{\left( \alpha - \beta \Lambda\right)}}$. Before analyzing these terms, note that, the $\ln$ function is negative when its argument is less than unity, positive when its argument is more than unity, and zero otherwise. Now, as the velocity increases from its initial value, that is when $u > u_0$, ($i$) becomes less than unity, and $(ii)$ becomes more than unity. However, as long as $u < \sqrt{\Lambda}$, the product of these terms remains less than unity and steadily decreases as $u$ increases. For this, the entire $\ln$ function takes the negative value, resulting in the continuous decrease in $x$, so is the winding point. This is exactly the reason why for $u < \sqrt{\Lambda}$ the velocity solution is upwinding as discussed in Section 3.1.1. 
At the terminus $\left (u = \sqrt{\Lambda}\right)$, $x$ is minimum with its negative value. After that, as soon as $u > \sqrt{\Lambda}$, the solution starts downwinding. 
Yet, it takes quite a while until the winding point comes back to the positive domain (or, $x \ge x_0$). How long it takes, depends on the value of ${\Lambda}$, including that of $\alpha, \beta, x_0$ and $u_0$, such that the condition $\displaystyle{\left (\frac{u_0}{u}\right)^{2 \beta \Lambda} \left( \frac{\alpha - \beta u_0^2}{\alpha - \beta u^2}\right)^{\left( \alpha - \beta \Lambda\right)} } > 1$ is satisfied. The higher the value of ${\Lambda}$ the farther the minimum of $x$ is pushed in the negative $x$ direction. So, the extended solution (\ref{Eqn_5}) has a lot of interesting and important physical aspects to explore. 
  
 \subsection{Essence of the new landslide velocity model} 
 
 Below, I illustrate some physical situations that are better explained by the new model than the existing ones.
 Figure \ref{Fig_1} shows that there are significant differences between the extended solution (\ref{Eqn_5}) and the simple solution (\ref{Eqn_6}). The difference widens strongly as $\Lambda$ increases. 
\begin{figure}[t!]
\begin{center}
 \includegraphics[width=15cm]{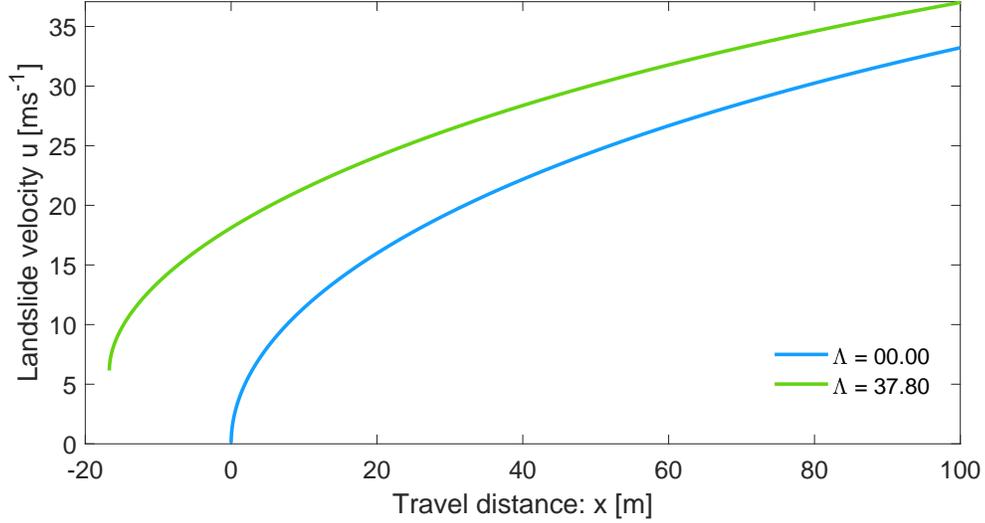}
  \end{center}
  \caption[]{Evolution of the landslide velocity as given by the extended landslide velocity model (\ref{Eqn_3}) and its analytical solution (\ref{Eqn_5}) for the initial conditions $(x_0, u_0) = (-16.70, 6.15)$, and $(x_0, u_0) = (0.0, 0.0)$. Only the downwind branch of the extended solution with $\Lambda = 37.80$ is considered. The left end of the extended solution can be extrapolated such that it hits at some location of the upstream sector of the channel, then practically this location may be taken as the landslide trigger point.}
  \label{Fig_2}
\end{figure}
 Comparison between the solutions (\ref{Eqn_5}) and (\ref{Eqn_6}) should be done carefully. One way to achieve this is by looking at the physically intuitive downwind dynamics starting from the terminus. This has been presented in Fig. \ref{Fig_2} for $\Lambda = 0$ and $\Lambda = 37.80$. The velocity solution that includes the force induced by the hydraulic pressure gradient (indicated by $\Lambda \neq 0$) and the depth of the landslide, which is physically important, is substantially higher than the one that neglects this force contribution. The difference is huge in the left and large in the right side of the figure. Even at $x = 100$ m, the difference between these solutions is 3.8 ms$^{-1}$. The difference increases steadily as the depth of the landslide increases. This is crystal clear from the physical perspective, because the higher depth of landslide induces the higher overburden load, that in turn increases the force induced by the hydraulic pressure gradient resulting in the enhanced motion. The new model and its solution explicitly demonstrate the essence of this basic physical fact which was ignored by all previous simple landslide models and their solutions. 
 \\[3mm]
 However, as seen in Fig. \ref{Fig_2}, the left end of the extended solution does not continuously decrease and hit the upstream sector of the slope. Yet, given the physical essence and superiority of the extended solution as compared with the simple solution, this problem can be technically eliminated. There are two possible ways. First, the left end point of the extended solution can be reasonably extrapolated such that it hits at some location of the upstream sector of the slope. Practically, this location can be taken as the triggering point of the landslide. Second, one might also seek to consider the velocity of the landslide at a slope position such that $u \ge \sqrt{\Lambda}$. Then, from that point, the extended solution can be applied customarily. 
 \\[3mm]
 \begin{figure}[t!]
\begin{center}
 \includegraphics[width=15cm]{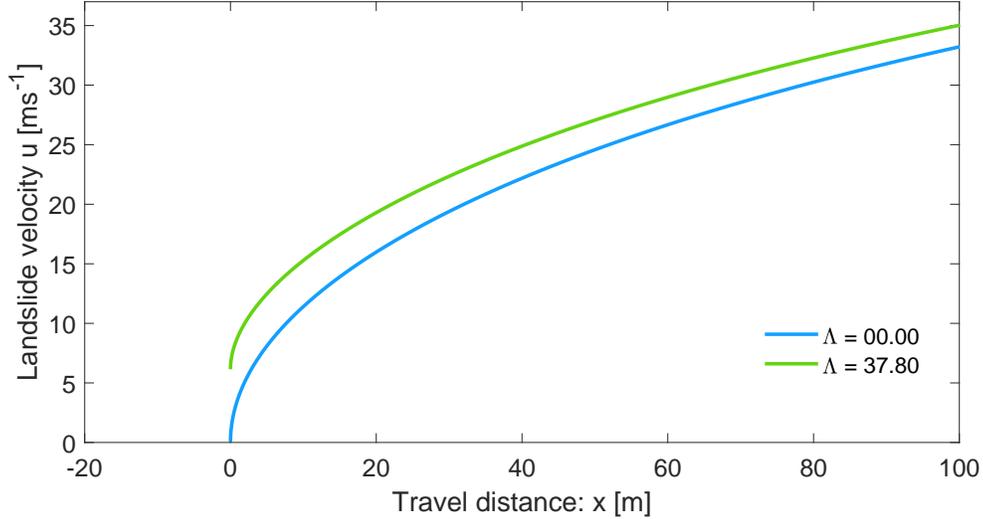}
  \end{center}
  \caption[]{Evolution of the landslide velocity as given by the extended landslide velocity model (\ref{Eqn_3}) and its analytical solution (\ref{Eqn_5}) for the transformed initial condition $(x_0, u_0) = (0.00, 6.15)$. Only the downwind branch of the extended solution is considered.}
  \label{Fig_2b}
\end{figure}
 Another probably practically more relevant comparison is presented in Fig. \ref{Fig_2b}, where the physically more pertinent downwind solutions are contrasted. This was made possible by transforming the general solution in Fig. \ref{Fig_2} (to the positive domain) with the initial condition $(x_0, u_0) = (0.00, 6.15)$, where the initial velocity is obtained from the relation $u = \sqrt{\Lambda}$. This allows a more closer look between the two solutions. However, note that, this transformation reduces the gap between the two solutions. Even for this, the difference in the velocity is significant. The difference is about 3.3 ms$^{-1}$ at position $x = 20$ m, which is about 22\% larger than the simple solution that disregards the force induced by the hydraulic pressure gradient that is present in landslide.
 \\[3mm]
 Except for some paradoxical situations (for $u < \Lambda$), solutions presented in Fig. \ref{Fig_1} $-$ Fig. \ref{Fig_2b} are deemed natural, and clearly manifest the importance of the inclusion of the dynamically important force induced by the hydraulic pressure gradient and the depth of the landslide. 
 Moreover, the difference between these solutions increases steadily with the increase in the landslide depth (because $\Lambda$ is proportional to the flow depth). This can also be explained. If we carefully look at the term associated with the hydraulic pressure gradient, it contains $\partial h/\partial x$ and the overburden pressure ($\lambda h$, per unit density) that includes the material depth, $h$. It is more convenient now to realize this in (\ref{Eqn_2c}) with the inertial term $-(\lambda h/u)\partial u/\partial x$ that explicitly accommodates the landslide depth $h$. It tells that, as $h$ increases the inertia decreases, resulting in the increase of the landslide velocity. This has exactly been demonstrated in Fig. \ref{Fig_1} $-$ Fig. \ref{Fig_2b} in their admissible domains.
  \\[3mm]
 Because, the landslide velocity plays a dominant role in estimating the impact force exerted by the landslide, which is proportional to the square of its velocity, the difference between the two velocities mentioned above, in fact, demonstrates the importance of the inclusion of the force induced by the hydraulic pressure gradient and the depth in determining the landslide velocity. However, all the existing analytical landslide velocity models ignore this important aspect, including the model by Pudasaini and Krautblatter (2022). This signifies the essence of the new landslide velocity model (\ref{Eqn_3}) and its solution (\ref{Eqn_5}).

\section{A simple mechanical model for drag}

\subsection{The drag model}

The viscous drag plays a very important role in controlling and describing the dynamics of mass flow (Pudasaini and Krautblatter, 2022). However, no clear and explicit mechanical and analytical model exists to date for the drag coefficient, and in practical and engineering applications, drag coefficients are exclusively numerically selected based on the model validation (Frank et al., 2015; Mergili et al., 2020; Shugar et al., 2021). Next, I utilize (\ref{Eqn_3}) and construct an analytical expression for the viscous drag coefficient $\beta$. 
The structure of (\ref{Eqn_3}) manifests that as the particle propagates closely with the wave speed $u \approx \sqrt{\lambda h}$, the force balance $\alpha - \beta u^2 \approx 0$ could be attained. This is a novel understanding resulting from the very graceful form of (\ref{Eqn_3}).
These two relations can be combined to eliminate $u$, and obtain a simple analytical, mechanical model for the viscous drag coefficient:
\begin{eqnarray}
{\displaystyle \beta \approx \frac{\alpha}{\lambda}\frac{1}{h}}, 
\label{Eqn_7}
\end{eqnarray}
where, as defined in (\ref{Eqn_2al}) and (\ref{Eqn_2la}), $\alpha$ is the net driving force (per unit mass) along the direction of motion composed of the gravitational acceleration and the frictional dissipation, and $\lambda$ is the coefficient of the hydraulic pressure gradient, that constitutes the lateral force (here, in the main downslope direction). Moreover, $h$ is the flow depth. 
Structurally, (\ref{Eqn_7}) is simple, yet, a mechanically extensive, innovative model. It is important to note that (\ref{Eqn_7}) is a dynamically evolving drag coefficient which may vanish as the net driving force nears zero, e.g., as the slope angle approaches the friction angle.
\\[3mm]
For dry and extensional ($K \approx 1$) landslides,  $\alpha $ and $\lambda$ are simplified. Then, for the slope angle $\zeta$ and the friction angle $\delta$, we have $\alpha = g\sin\zeta - g\tan\delta\cos\zeta $, $\lambda = g\cos\zeta$. For this, (\ref{Eqn_7}) reduces to 
\begin{eqnarray}
{\displaystyle \beta = \frac{1}{h}}\tan\zeta\left [ 1- \frac{\tan\delta}{\tan\zeta}\right ].
\label{Eqn_8}
\end{eqnarray}
There is nothing unknown in (\ref{Eqn_7}) or (\ref{Eqn_8}), because the flow depth $h$ can either be assumed (known), or be considered as a dynamical quantity, that is already in the balance equations, and all other physical, mechanical and geometrical parameters in $\alpha$ and $\lambda$ are known or given. So, (\ref{Eqn_7}) and (\ref{Eqn_8}) are closed, and can directly be applied (in the momentum equation) in modeling and simulation of mass transport.

\subsection{Importance of the new drag model}

The more general drag coefficient (\ref{Eqn_7}) should be used in modelling mass flows by replacing $\beta$ in the momentum balance equation. There are several important aspects associated with (\ref{Eqn_7}). 

\begin{itemize}

\item [($i$)] 
As all the involved physical parameters are known, (\ref{Eqn_7}) offers a new physical-mathematical model for the drag coefficient for mass flow simulation, covering dense to dilute flows (due to the presence of the solid fraction, $\alpha_s$, and the lubrication factor, $1-\gamma$, in $\alpha$ and $\lambda$). 
The Manning-Strickler formula (Ackers and White, 1973) for open channel hydraulics for the friction coefficient explicitly depends inversely on the one-third of the flow depth (i.e., $1/h^{1/3}$). However, the drag coefficient in (\ref{Eqn_7}) depends inversely on the flow depth (i.e., $1/h$). 
Analogous to Manning's or Strickler's number, I call ${\alpha}/{\lambda}$ the drag number (parameter), and write as $\displaystyle{\mathcal P_d = {\alpha}/{\lambda}}$, which is a dimensionless number, as a ratio of two forces; the net driving force along the slope and the lateral force associated with the hydraulic pressure gradient. Thus, the drag coefficient $\beta$ is the ratio between the drag number $\mathcal P_d$ and the flow depth $h$: $\beta = \mathcal P_d/h$.

 \item [($ii$)] The drag coefficient increases with decreasing flow height. This means, thin flows experience higher drag which is in agreement with the fluid-type drag often used in simulation of mass flows (Mergili et al., 2020) to acquire observed results. But now, I have presented a physically-founded analytical drag model that serves the purpose of practitioners and engineers. 

\item [($iii$)] Drag increases with increasing slope angle and decreases with increasing basal friction angle. This is consistent with the flow dynamics, because the higher slope and lower friction result in higher motion, which in turn, increases the drag force. In other words, the lower slope angle and higher friction reduces motion and the drag too. 

\item [($iv$)] As the slope angle decreases and approaches the friction angle (equivalently, as the friction angle approaches the slope angle), both the net driving force $\alpha$ and the viscous drag coefficient $\beta$ tend to vanish. This is the transition from motion to deposition or the flow stopping. This is how collectively and synchronously the entire system force ceases and the mass comes to a halt. 

\item [($v$)] The drag is also influenced by the particle concentration and density ratio between the fluid and the solid in the mixture as well as the extensional and compactional motions. 

\end{itemize}
For the first time, these realistic dynamically evolving situations could be described by my modelling approach which was not possible by any existing drag models as all those models use empirical drag coefficients. 

\subsection{Dynamics of the drag coefficient: the drag curves}

Here, I show, how the dynamics of drag coefficient $\beta$ are influenced by the flow depth ($h$), the slope angle ($\zeta$), the friction angle ($\delta$), and the earth pressure coefficient ($K$) for a dry landslide. These are called the drag curves.

\subsubsection{Influence of the flow depth}

\begin{figure}
\begin{center}
 \includegraphics[width=15cm]{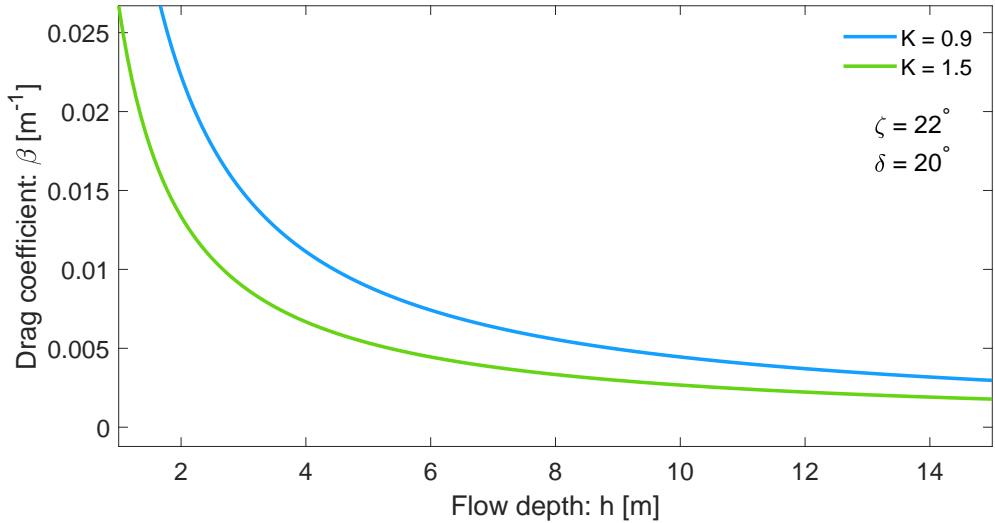}
  \end{center}
  \caption[]{The dynamics of the drag coefficient $\beta$ as a function of the flow depth $h$ showing the exponential increase of the drag coefficient as the flow depth decreases. Also shown are the physical parameters, the slope angle $\zeta$, the basal friction angle $\delta$, and the earth pressure coefficient $K$.}
  \label{Fig_3}
\end{figure}
Figure \ref{Fig_3} shows the dynamics of the viscous drag coefficient $\beta$ as a function of the flow depth $h$ for given parameters for the slope angle, friction angle, and the earth pressure coefficient, representing a landslide. This figure shows that the drag coefficient increases rapidly as the flow depth decreases, which is consistent with our intuition and the practical experience. Interestingly, as the flow depth becomes very high, the drop in the drag is very small. This means, in practice, we may set a realistic lower limit to the drag coefficient. The influence of the earth pressure coefficient $K$ on the drag is discussed separately in Section 4.3.4.

\subsubsection{Influence of the friction angle}

\begin{figure}[t!]
\begin{center}
 \includegraphics[width=15cm]{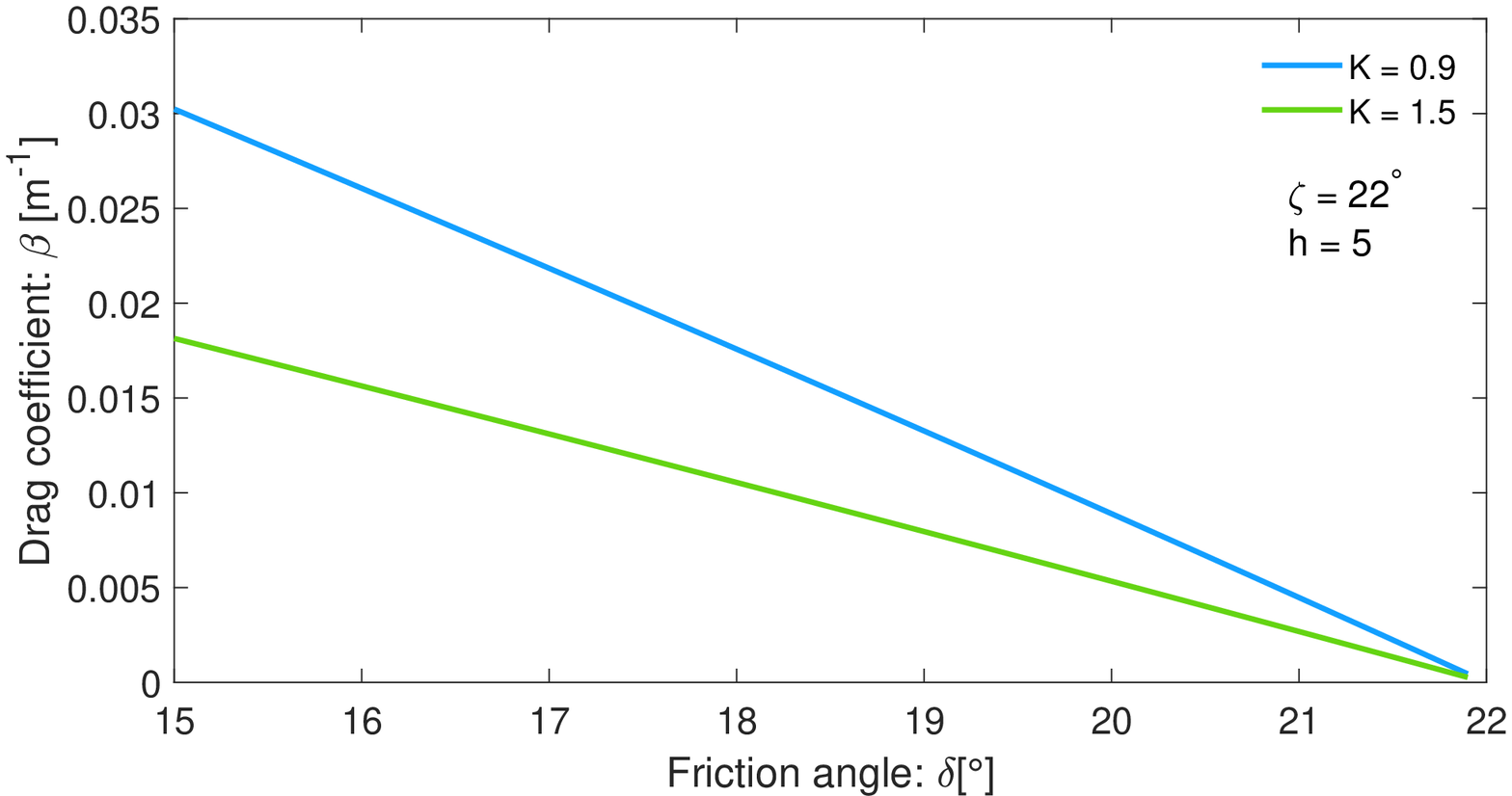}
  \end{center}
  \caption[]{The dynamics of the drag coefficient $\beta$ as a function of the basal friction angle $\delta$ indicating the almost linear decrease of the drag coefficient as the friction angle increases. Also shown are the physical parameters, the slope angle $\zeta$, the landslide depth $h$, and the earth pressure coefficient $K$.}
  \label{Fig_4}
\begin{center}
 \includegraphics[width=15cm]{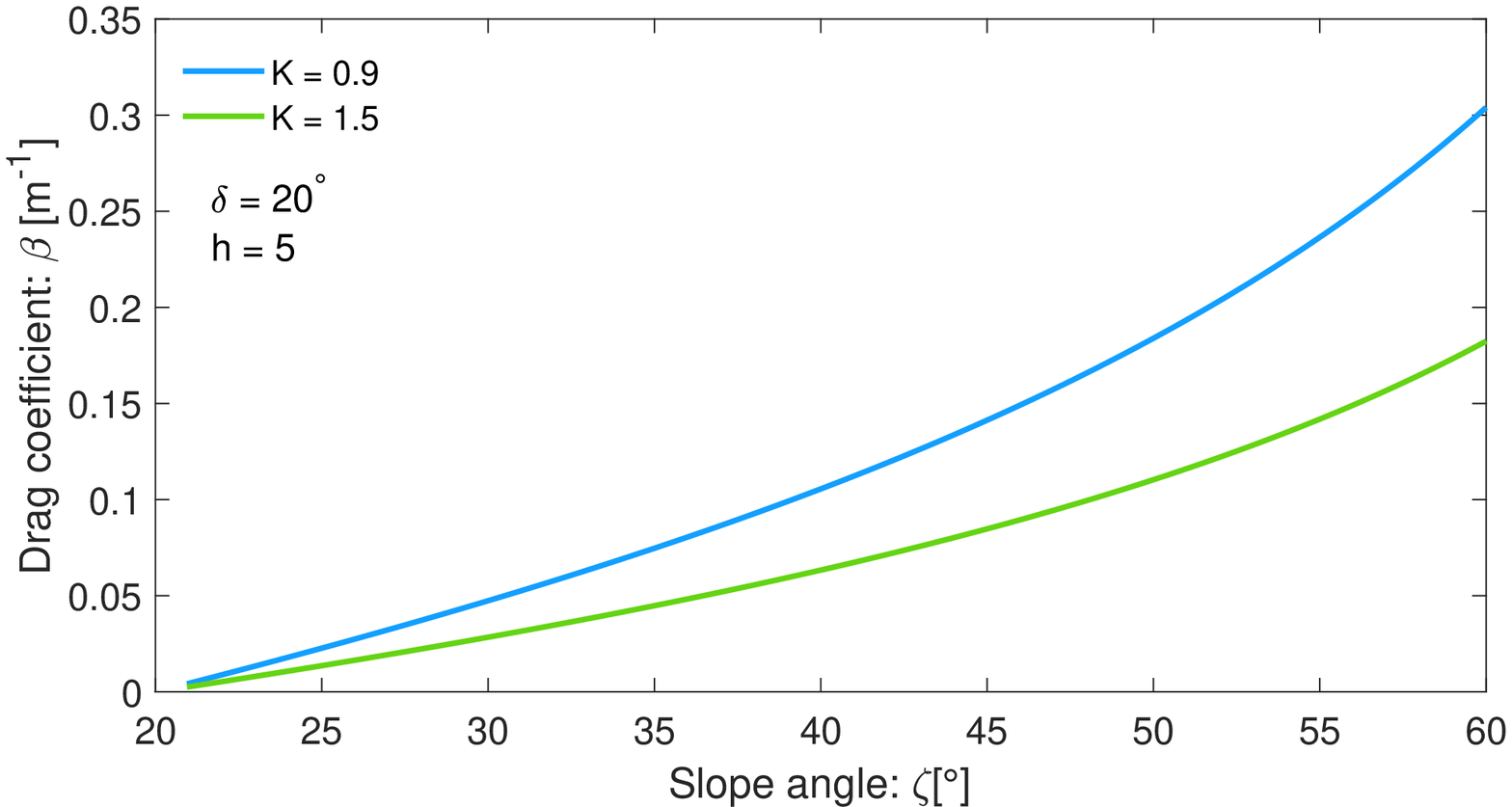}
  \end{center}
  \caption[]{The dynamics of the drag coefficient $\beta$ as a function of the slope angle $\zeta$. The drag coefficient increases non-linearly as the slope angle increases. Also shown are the physical parameters, the basal friction angle $\delta$, the landslide depth $h$, and the earth pressure coefficient $K$.}
  \label{Fig_5}
\end{figure}
Figure \ref{Fig_4} displays the dynamics of the viscous drag coefficient $\beta$ as a function of the basal friction angle $\delta$ for given parameters for a landslide. The drag coefficient decreases almost linearly as the basal friction angle increases. Interestingly, $\beta$ is eminently small as the fiction angle approaches the slope angle. This is as expected, because, as explained above, this is the state closer to the flow halting. 

\subsubsection{Influence of the slope angle}

The dynamics of the viscous drag coefficient $\beta$ as a function of the slope angle $\zeta$ for considered parameters is presented in Figure \ref{Fig_5}. Now, the drag coefficient increases non-linearly as the slope angle increases. The drag coefficient is nominally small as the slope angle reduces significantly and approaches the basal friction angle. This is the near cease situation of motion. 

\subsubsection{Expanding and contracting motions}

As it slides down, usually the landslide accelerates, it expands and the corresponding earth pressure coefficient $K$ takes the values in the range $K \le 1$. However, as it enters the transition zone, propagates in the run-out area, and begins deposition, the motion switches from accelerating to decelerating states. Consequently, the landslide mass contracts. In this situation, the earth pressure coefficient can become much higher than unity.
The earth pressure coefficient for the expanding or contracting motions can be determined dynamically as (Pudasaini and Hutter, 2007):
\begin{eqnarray}
{\displaystyle 
K = 2\sec^2\phi\left( 1 \mp \sqrt{1-\cos^2\phi\sec^2\delta}\right ) -1
},
\label{Eqn_K}
\end{eqnarray} 
where, $\phi$ is the friction angle between the grains, the $-$ sign corresponds to the expanding motion ($\partial u/\partial x \ge 0$), and the $+$ sign corresponds to the contracting motion ($\partial u/\partial x < 0$).
For the substantially expanding and contracting motions, I respectively took $K = 0.9$, and $K = 1.5$.
\\[3mm]
The results in Fig. \ref{Fig_3} $-$ Fig. \ref{Fig_5} reveal that, as landslide changes its state from extensional to compactional motion, the drag coefficient $\beta$ behaves completely differently. It decreases substantially with increasing values of $K$. 
This is physically meaningful, because as the mass starts depositing, being in the state of contraction, naturally, the landslide should experience less drag than in extending region. Otherwise, the process of mass deceleration and transition in to deposition is unphysically rapid if the higher drag is applied as in the accelerating region, resulting in unrealistically reduced runout. This means, drag coefficients are essentially different for expanding and contracting motions. This is an entirely novel understanding. In applications, often, uniform numerical drag coefficient is applied globally for the entire flow path $-$ from triggering to the main track to transition to run-out and deposition without making any distinction between the extensional and compactional motions. The present analysis proves that this is unphysical. So, different drag coefficients must be applied consistently in the mass expanding and contracting regions. These regions are dynamically determined by the accelerating and decelerating motions and their corresponding $K$ values, less than unity or greater than unity (Pudasaini and Hutter, 2007).
\\[3mm]
Importantly, the values of the drag coefficients displayed in Fig. \ref{Fig_3} $-$ Fig. \ref{Fig_5} are close to the empirical or the numerical drag coefficients often used in practice in mass flow simulations (Mergili et al., 2020; Shugar et al., 2021). But, previously, those values were taken without any physical foundation. We have now the physically described, fully mechanical and analytical drag model that can be instantly applied in mass flow simulations.

\section{Summary}

The landslide velocity plays a decisive role in estimating the impact force and havoc. Here, I developed a novel extended landslide velocity model that includes the force induced by the hydraulic pressure gradient, which, however, was neglected by all the existing analytical landslide velocity models. By a rigorous and fully consistent conversion between this force and inertia, I developed two peer systems that are expected to produce the same results. However, this contradicts with our conventional wisdom that the mass must accelerate downslope. This is a new understanding. The solutions produced by these two systems are fundamentally different. As combining the momentum balance with the mass balance is a common practice in fluid mechanics, I have proven that, one must very carefully analyze the dynamics of the resulting system, and check its functionality. 
\\[3mm]
The new extended landslide velocity model is simple, elegant, and yet an extraordinary development that characterizes some key features of the flow dynamics, and leads to the pressure-inertia-paradox. It describes some unexpected, but important situations emanating for the first time, the upwind and downwind solutions. The mass first moves upstream, then it winds back and accelerates down slope. This feature intensifies rapidly as the flow depth increases. The mechanism of the new solution is explained based on the contribution from the hydraulic pressure gradient. I compare the two landslide velocity models that neglects and includes the force induced by the hydraulic pressure gradient. There are significant to huge differences between the extended and the simple solutions. The difference widens strongly as the lateral force associated with the hydraulic pressure gradient increases. This signifies the essence of the extended landslide velocity model and its solution.
\\[3mm]
The viscous drag plays an important role in controlling the dynamics of mass flow. However, no explicit mechanical and analytical model exists to date for drag. In practice, drag coefficients are exclusively numerically selected. Here, the very graceful form of the new velocity model provided the unique opportunity to construct a simple, mechanically extensive, analytical model for the viscous drag. It offers an innovative, physically-founded model for the drag, covering dense to dilute flows. It varies inversely with the flow depth and contains several physical, mechanical and geometrical parameters associated with the flow. The model features that as both the net driving force and the viscous drag coefficient tend to vanish, concomitantly the entire system force ceases resulting in the mass halting. I constructed a dimensionless drag number as the ratio between the net driving force and the lateral force. Importance of the drag model and its practical applicability is explained. The drag coefficient increases rapidly as the flow depth decreases, it decreases almost linearly as the basal friction angle increases, and it increases non-linearly as the slope angle increases. This is consistent with our experience. In contrast to the prevailing practices, where drags are chosen uniformly for the entire flow path, I have proven that drags are essentially different for the expanding and contracting motions. So, different drags must be applied consistently in different regions of the landslide. This is an entirely novel perception. The drag coefficients are close to the empirical or the numerical values often used in practice. So, the new drag model serves the purpose of practitioners and engineers dealing with mass flow simulations.


 \end{document}